\documentclass[preprint2]{aastex}

\shorttitle{Statistics of SNRs} \shortauthors{J. W. Xu \& F. J. Lu}

\begin{document}

\title{Statistics of Galactic Supernova Remnants (continued)}

\author{Jian-Wen Xu\altaffilmark{1} and Ya-Peng Hu\altaffilmark{2}}
\affil{Key Laboratory of Frontiers in Theoretical Physics, Institute
of Theoretical Physics, Chinese Academy of Sciences, Beijing 100080,
China}

\altaffiltext{1}{Postdoctor, Institute of Theoretical Physics,
Chinese Academy of Sciences, Beijing 100080, China.}
\altaffiltext{2}{PhD candidate, Institute of Theoretical Physics,
Chinese Academy of Sciences, Beijing 100080, China.}
\email{xjw@itp.ac.cn}

\begin{abstract}
Our statistics on Galactic supernova remnants (SNRs) shows that the
electrons temperature ($T$) of hard X-ray and the shock waves
traveling velocity ($\upsilon$) decreases with ages ($t$) for
all-sort remnants. However, the shock waves swept-up mass ($M_{su}$)
of ISM increases with the age. Second, the remnant radio fluxes
($S$) at 1~GHz increase slightly with ISM electrons density ($n_0$).
At last, the number distributions illustrate that the supernovae
(SNe) initial kinetic energy ($E_0$), hydrogen column density
($N_H$), electrons temperature (kT) of hard X-ray, magnetic field
($B$) and the shock waves swept-up mass ($M_{su}$) of ISM mainly
peaked at $(1 \sim 10) \times 10^{50}$~ergs, $(1 \sim 10)\times
10^{21}$~cm$^{-2}$, a few KeV, 100~$\mu$G and
10$\sim$100~$M_{\odot}$, respectively.
\end{abstract}


\section{Introduction}
\label{sec:intro}

The statistics on supernova remnants physical parameter have been
made by many authors before (eg. Poveda \& Woltjer 1968; Clark \&
Caswell 1976; Lozinskaya 1981; Green 1984; Mills et al. 1984;
Allakhverdiyev et al. 1985; Huang \& Thaddeus 1985; Duric \&
Seaquist 1986; Guseinov et al. 2003; Arbutina et al. 2004). Most of
them made about the relation between remnants radio surface
brightness ($\Sigma$) and their diameter ($D$). Xu et al. (2005) had
done a statistics of Galactic SNRs in more details with larger
candidate samples. However statistics on other parameters, i.e. the
shell swept-up mass ($M_{su}$), electrons temperature ($T$) of hard
X-rays, remnants magnetic field ($B$), shock wave traveling velocity
($\upsilon$) and electron density ($n_0$) of the interstellar media
(ISM) have rarely being done even since.

Here we collected most of Galactic remnants parameters
(table~\ref{tab1}) mentioned above to make a more detail statistics
trying to find more physical connections, which also as a continued
work of Xu et al. (2005). We concisely explain how to derive all
these remnant parameters in Sect.~\ref{sec:parameter}, offer the
statistical results in Sect.~\ref{sec:statis}. And at last summarize
our conclusion.

\section{Parameters of SNRs}
\label{sec:parameter}

Former paper (Xu et al. 2005) has described how some basic physical
parameters of Galactic SNRs, i.e. the evolved ages ($t$), kinematic
distances ($d$), spectral index ($\alpha$) and the vertical height
($z$) from the galactic plane etc., were estimated. Here more
remnants parameters are added up. Many of the radio SNRs own more
than one published parameter values. We either chose the most recent
estimated ones or adopted an average of the available estimates, or
the most commonly used value.

Now let us concisely explain how some basic physical parameters of
the supernova remnants are derived, since our statistical work would
be based on these.

\subsection{Age Estimates}

\clearpage

\begin{figure*}
\begin{tabular}{ccc}
\includegraphics[bb=155 45 728 707,width=0.4\textwidth,angle=-90]{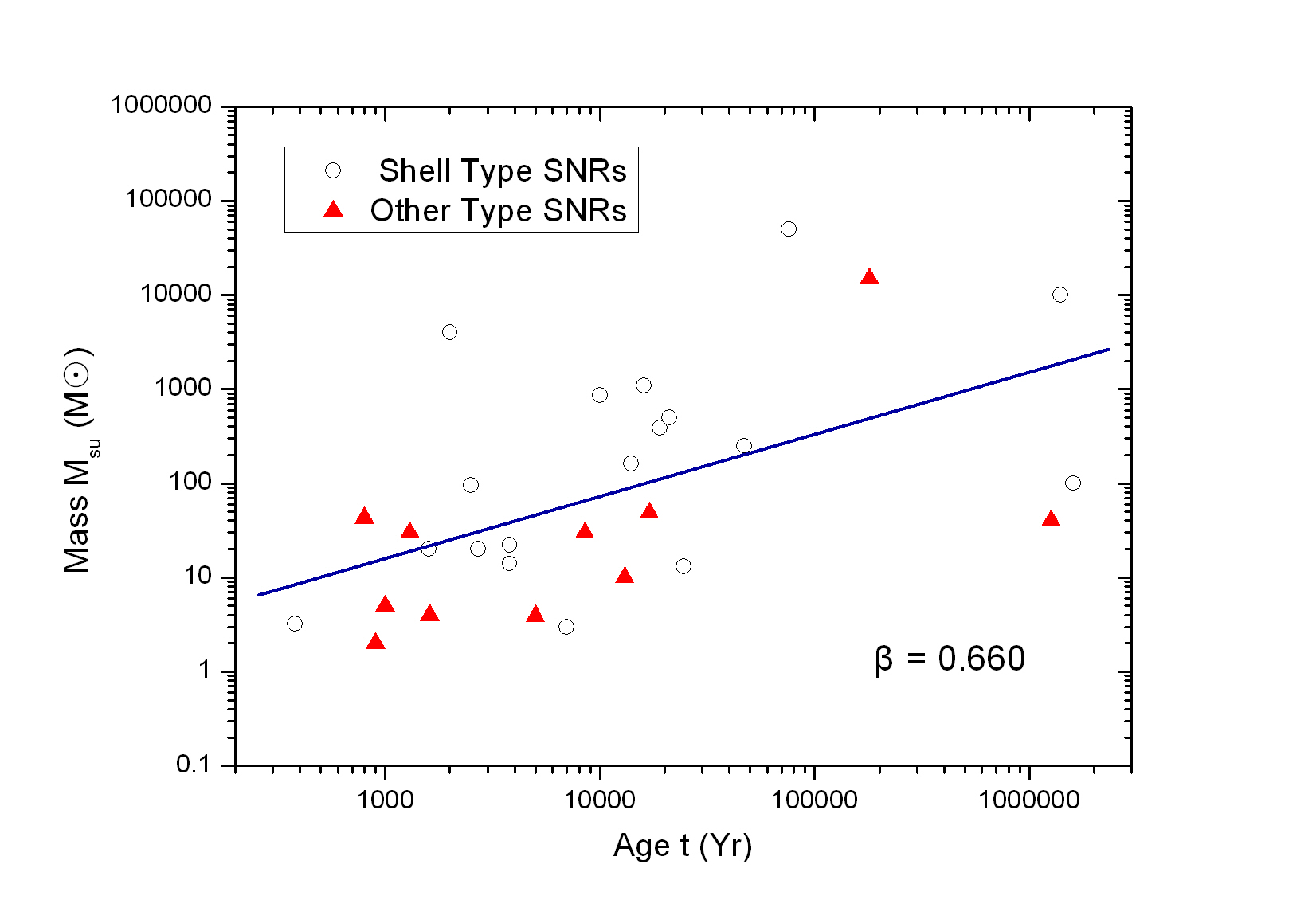}&
\includegraphics[bb=153 41 726 709,width=0.4\textwidth,angle=-90]{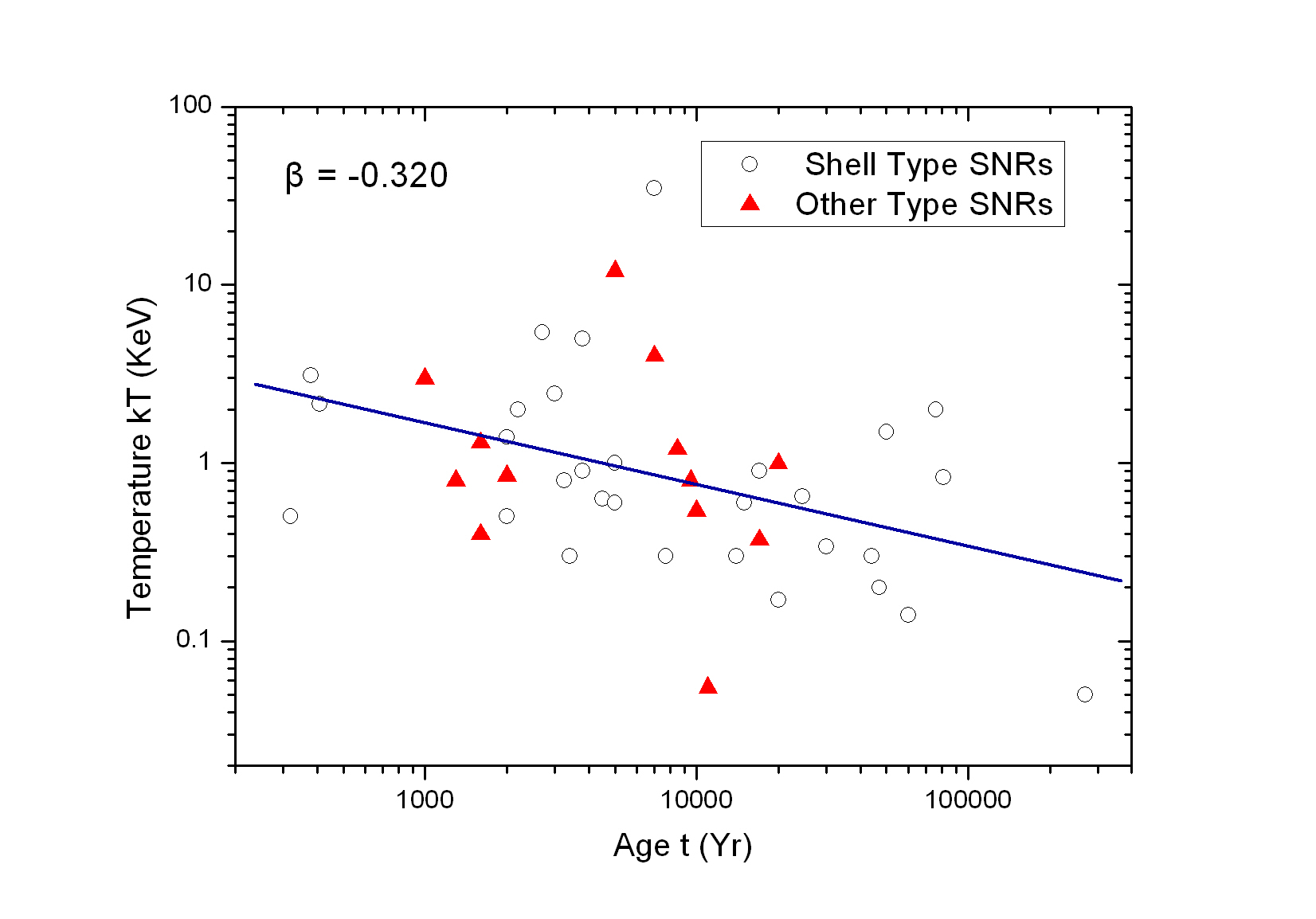}\\
\includegraphics[bb=151 45 730 707,width=0.4\textwidth,angle=-90]{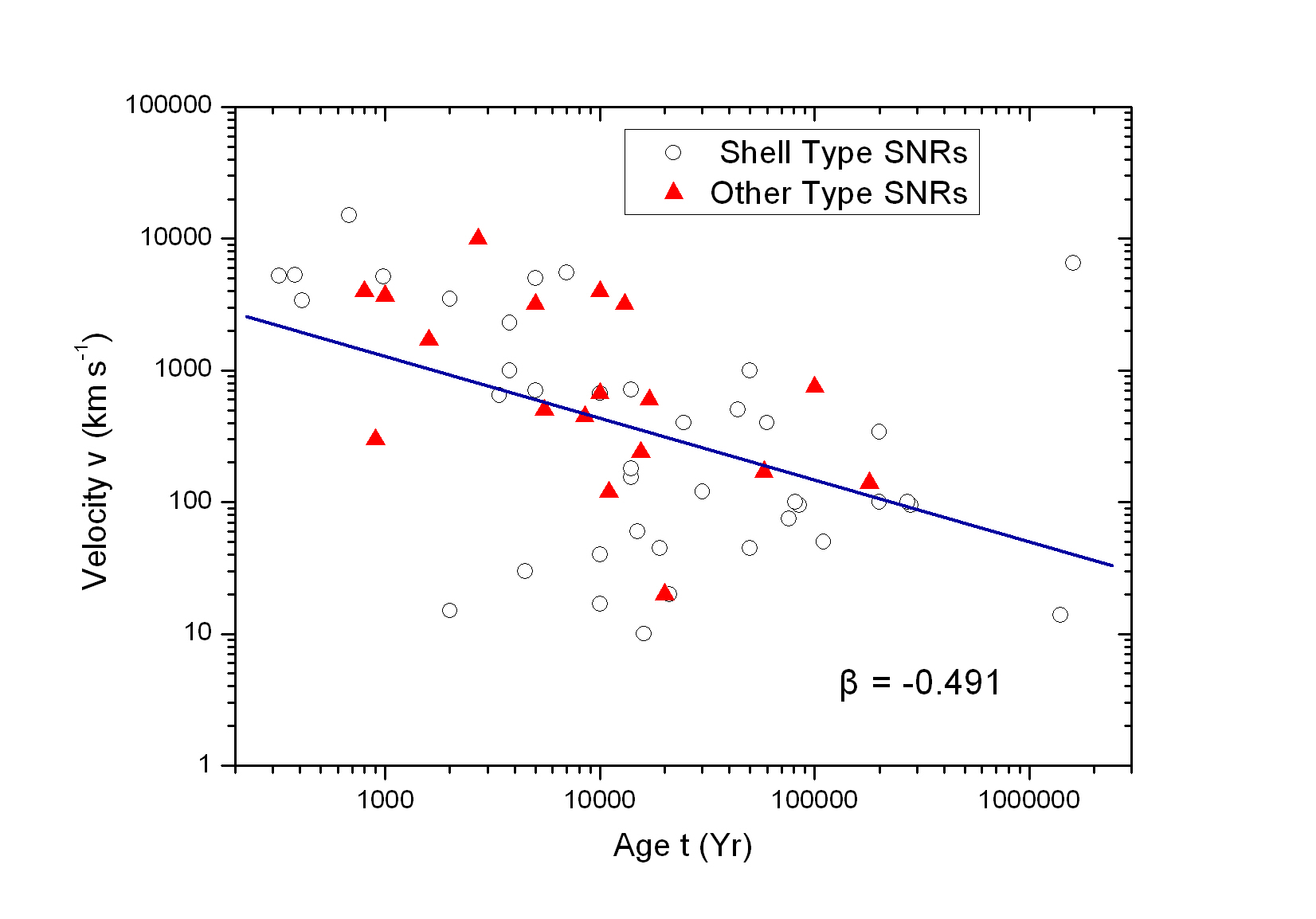}&
\includegraphics[bb=155 45 728 707,width=0.4\textwidth,angle=-90]{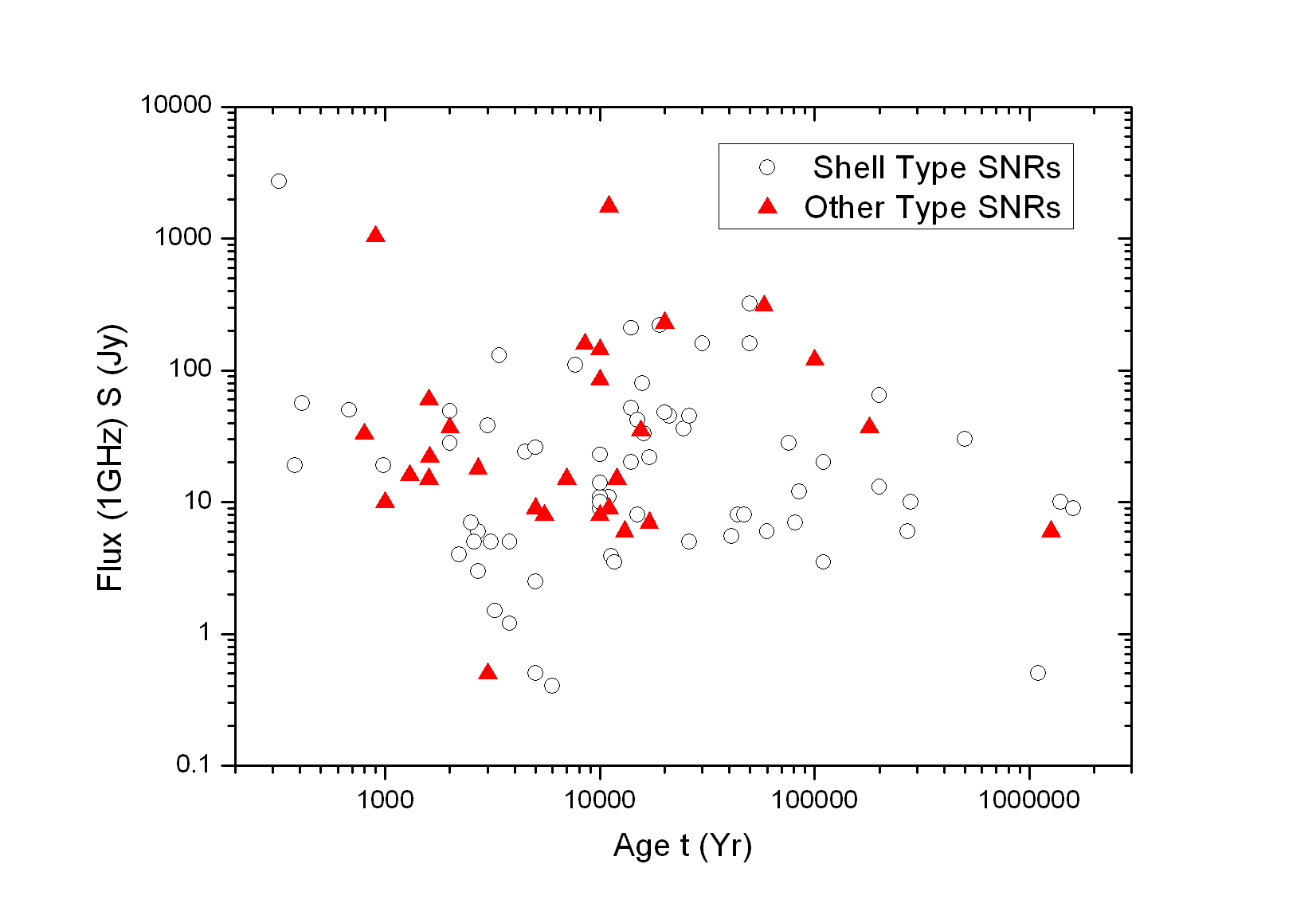}\\
\includegraphics[bb=151 45 730 707,width=0.4\textwidth,angle=-90]{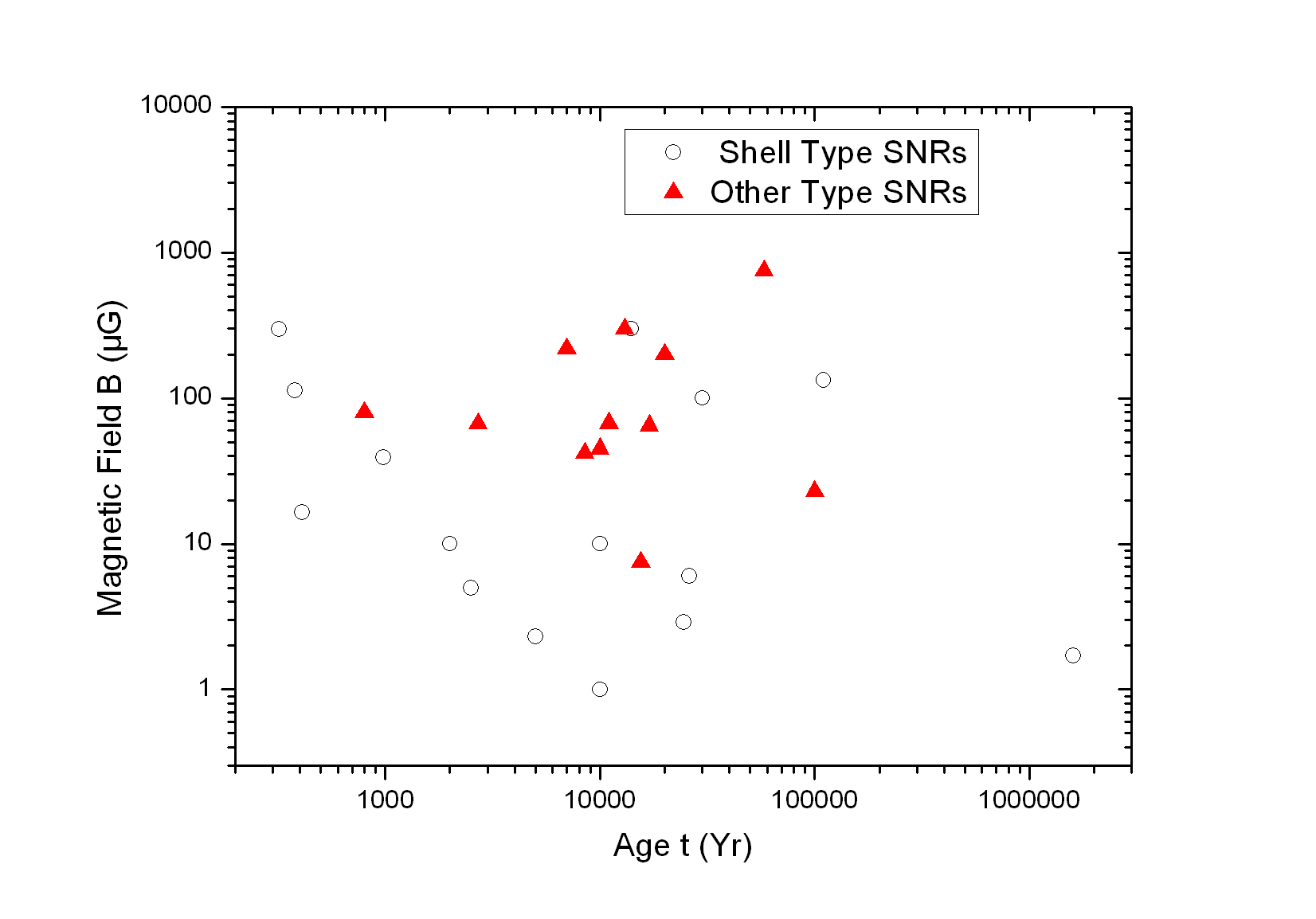}&
\includegraphics[bb=153 41 726 709,width=0.4\textwidth,angle=-90]{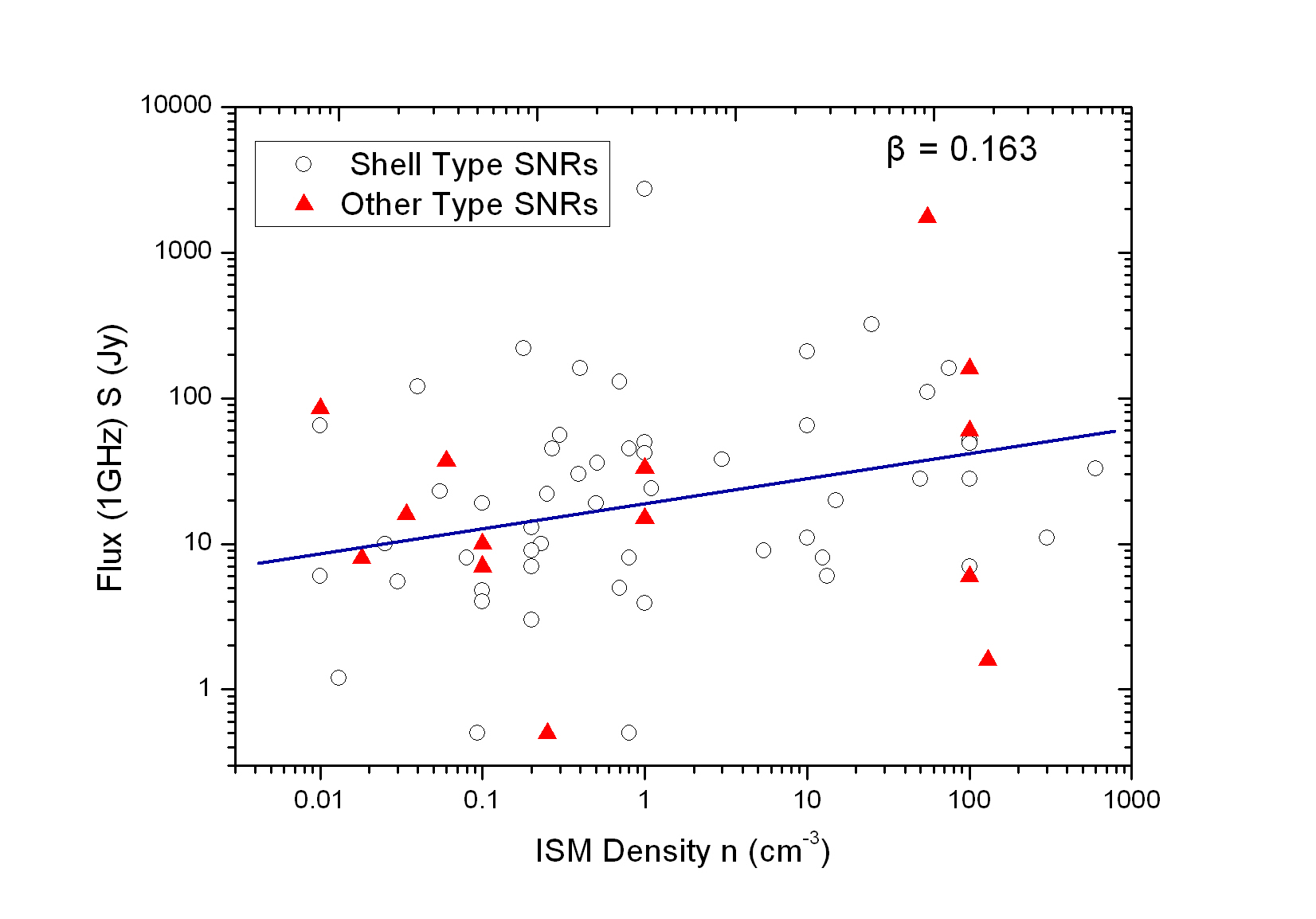}\\
\end{tabular}
\caption{Images show the correlations of the physical parameters of
both shell-type and other type SNRs in Galaxy. Some of them are
labeled with their index ($\beta$) of the best fitting line. Here $Y
\propto X^{\beta}$, $X$ is the x-axis variable, and $Y$ the y-axis
one.}\label{fig:rela}
\end{figure*}

\clearpage

\begin{figure*}
\begin{tabular}{ccc}
\includegraphics[bb=55 35 708 607,width=0.4\textwidth,angle=0]{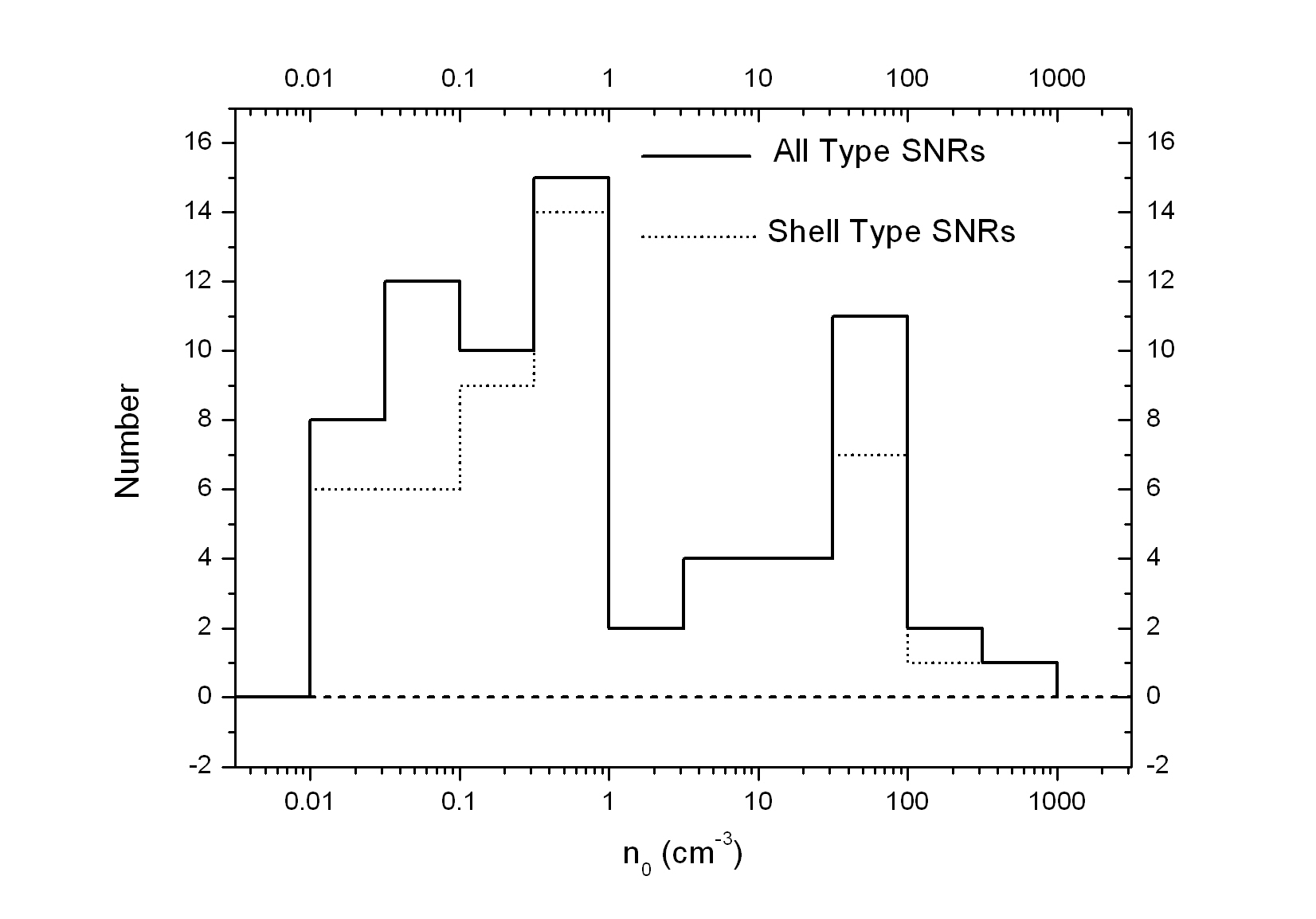}&
\includegraphics[bb=51 35 700 607,width=0.4\textwidth,angle=0]{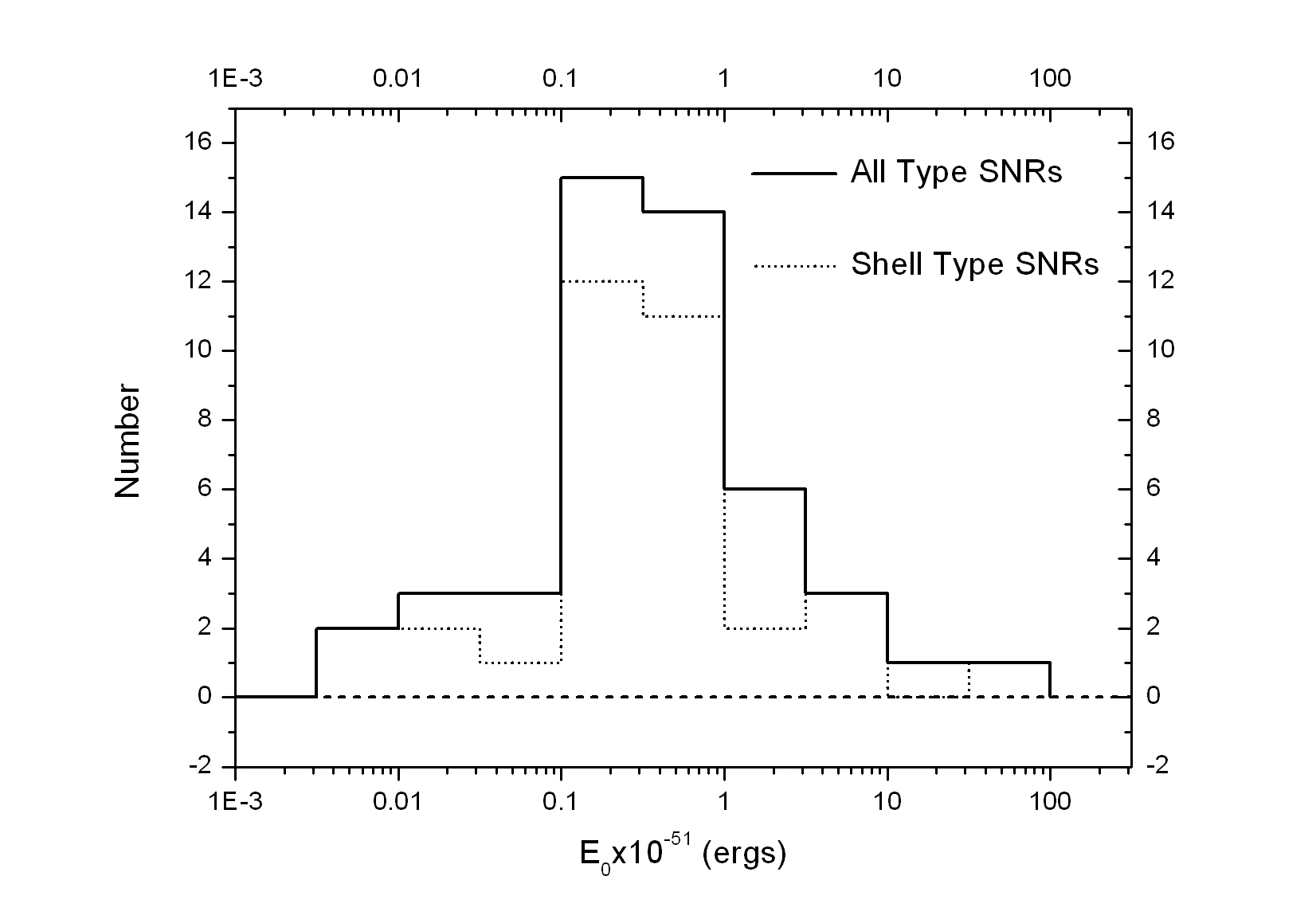}\\
\includegraphics[bb=53 31 706 609,width=0.4\textwidth,angle=0]{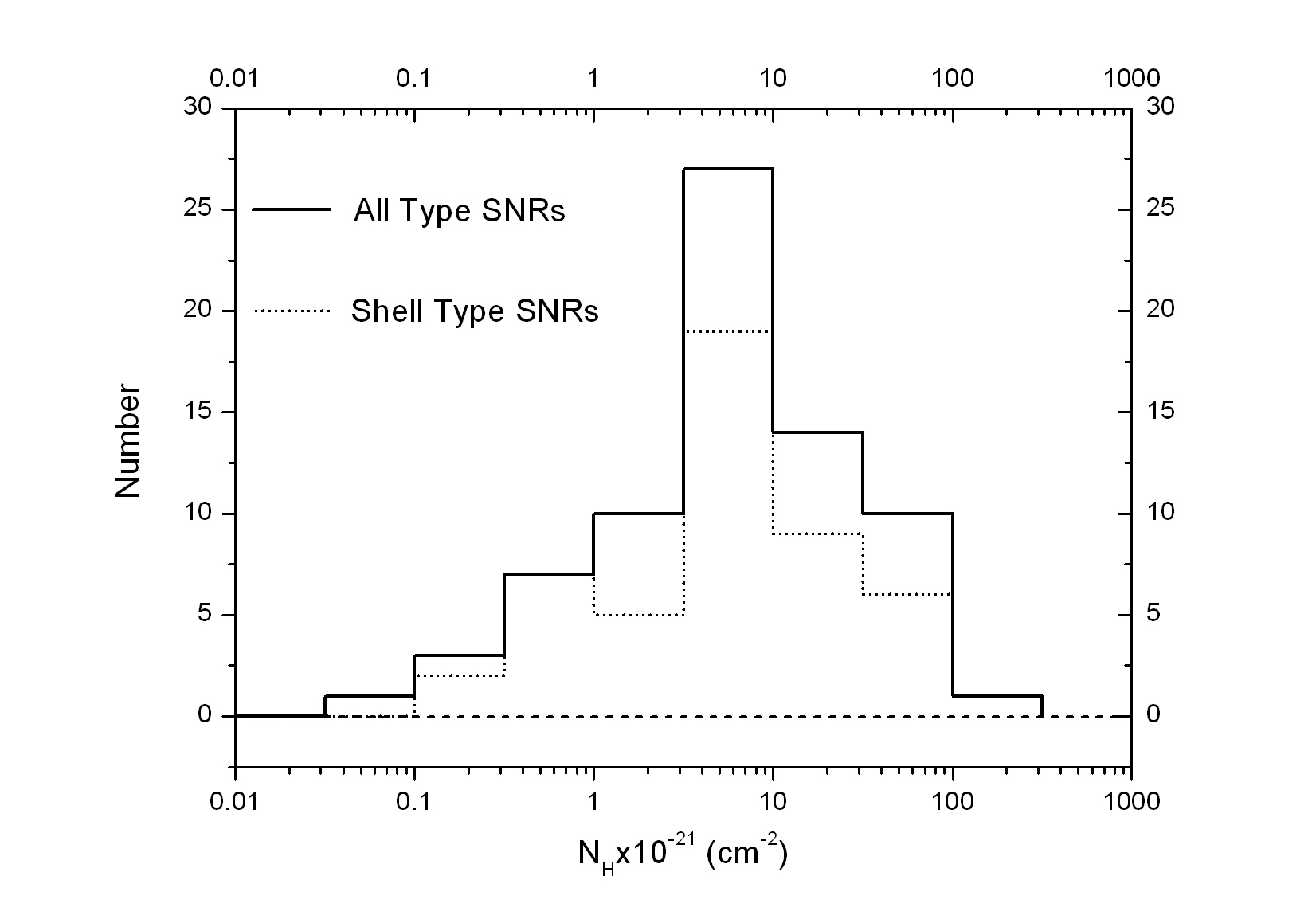}&
\includegraphics[bb=55 35 708 607,width=0.4\textwidth,angle=0]{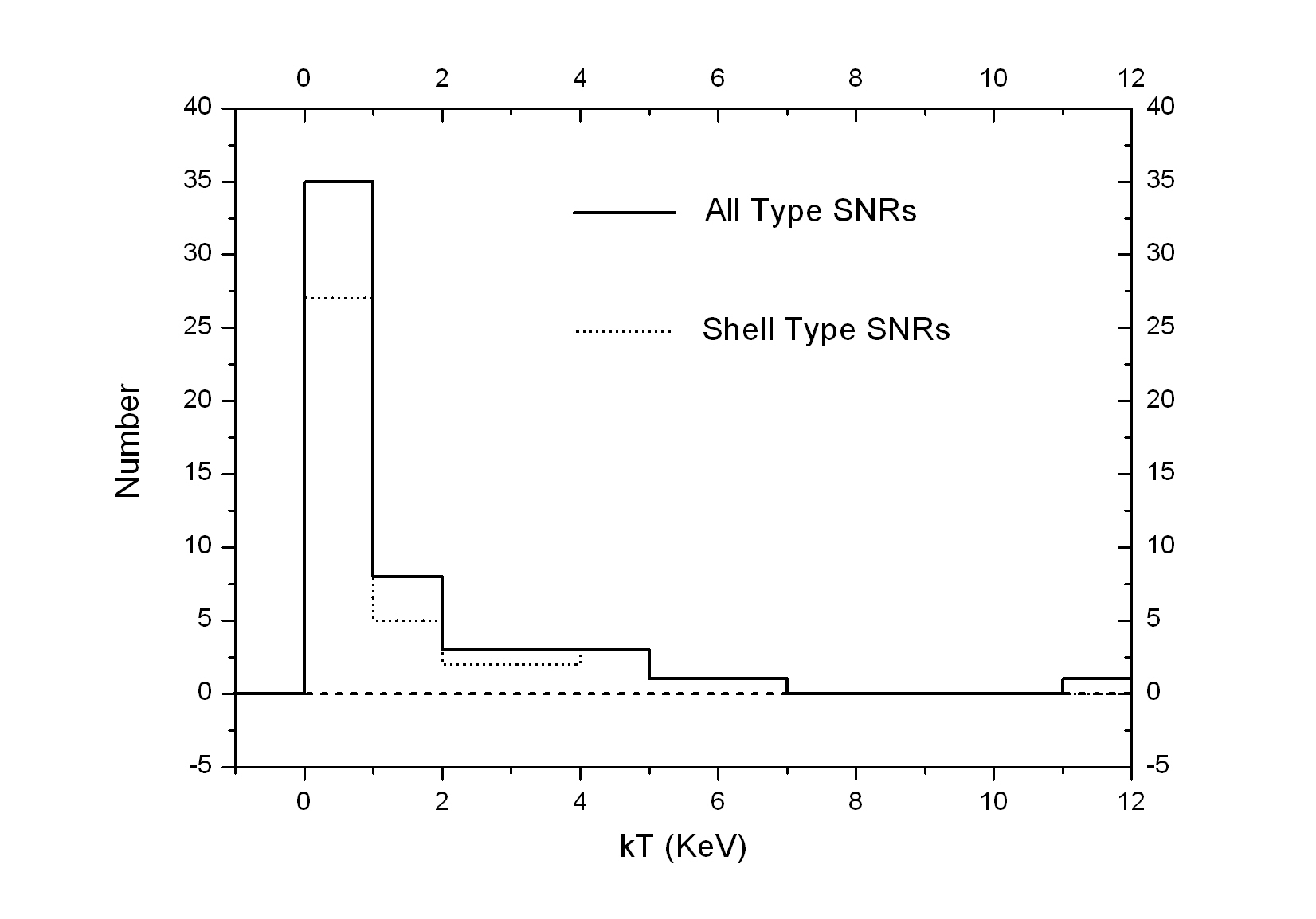}\\
\includegraphics[bb=51 35 700 607,width=0.4\textwidth,angle=0]{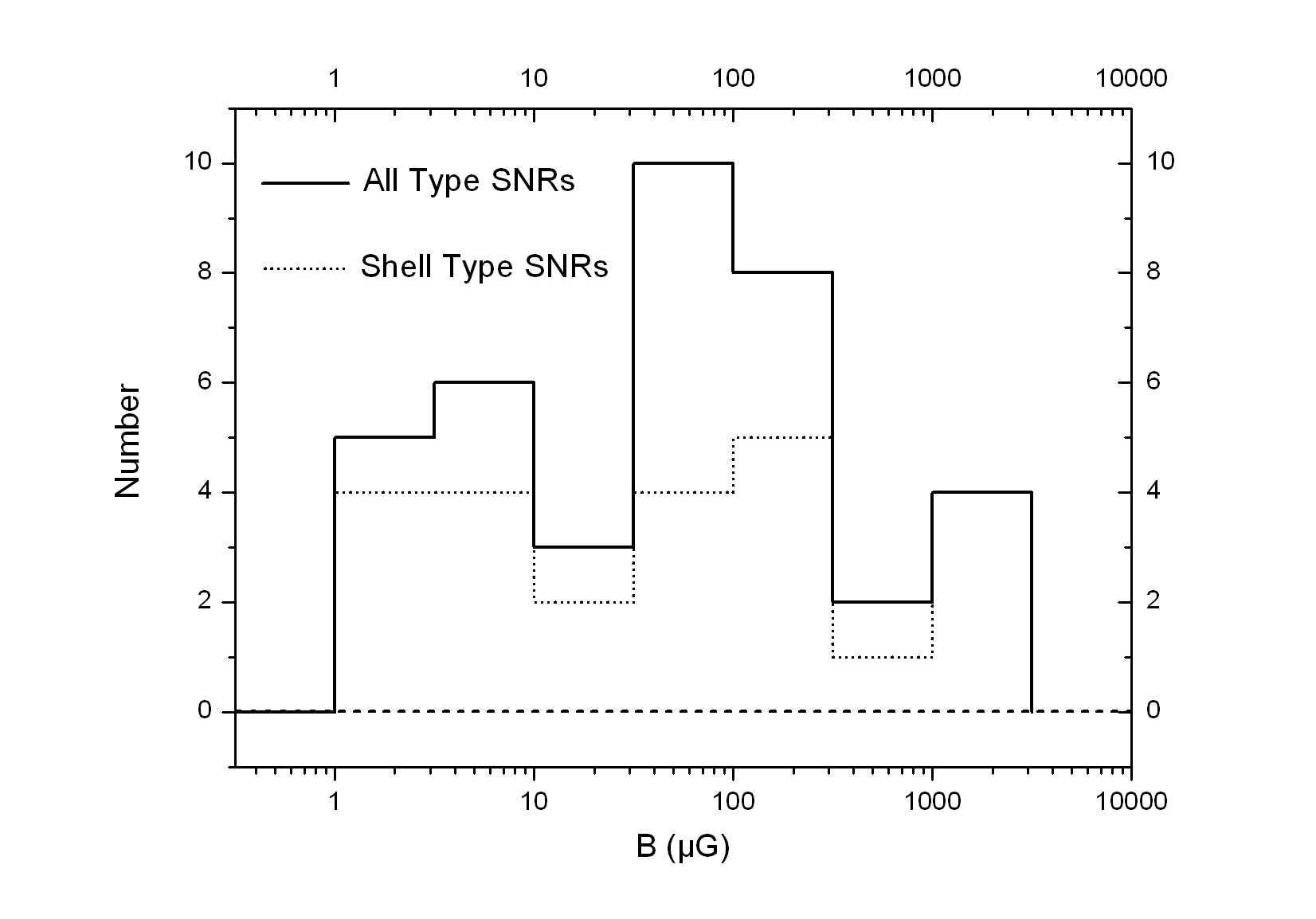}&
\includegraphics[bb=53 31 706 609,width=0.4\textwidth,angle=0]{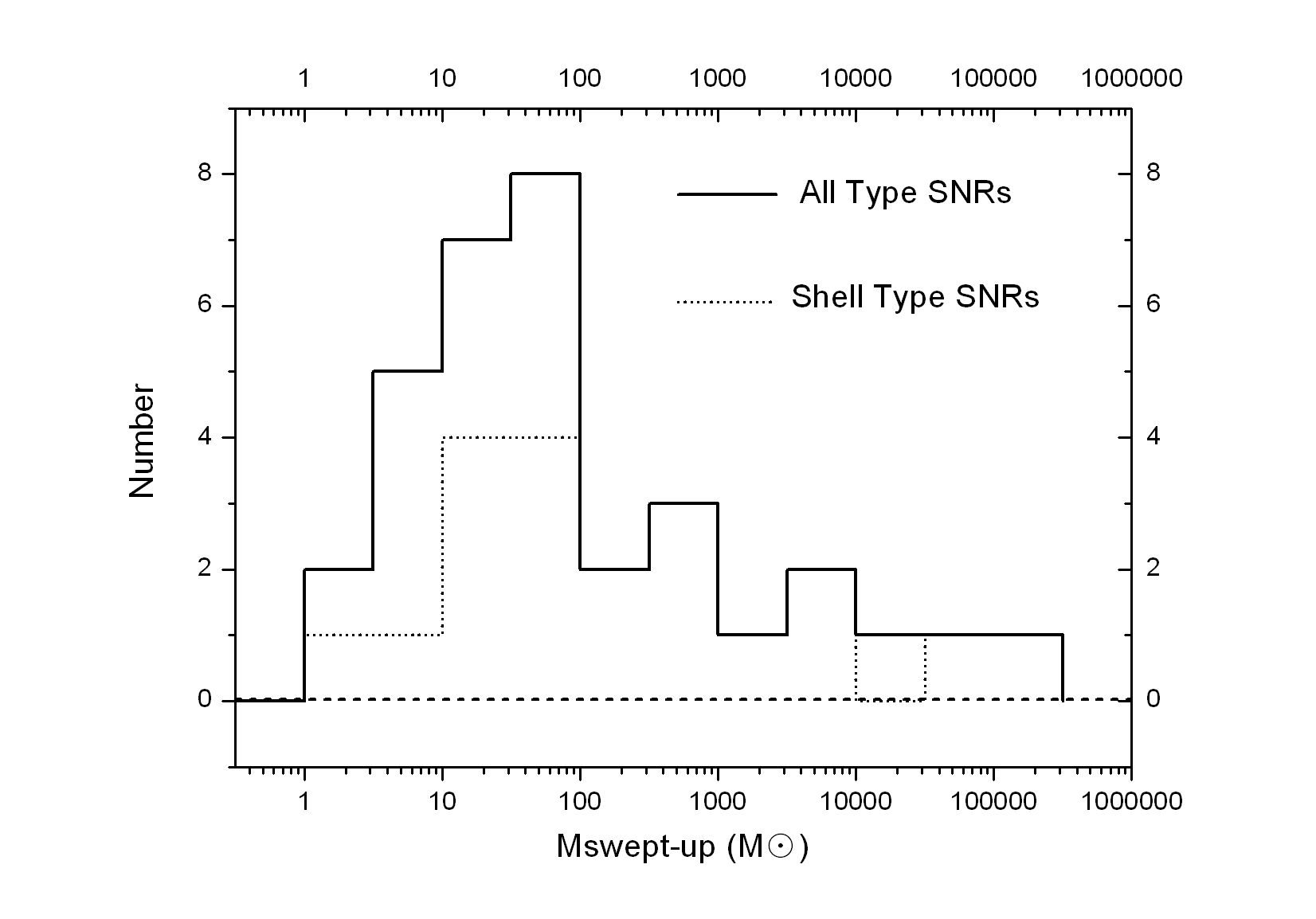}\\
\end{tabular}
\caption{Diagrams denote the number distributions of the electron
density ($n_0$), SNe initial explosion energy ($E_0$), hydrogen
column density ($N_H$), electron temperature (kT), magnetic field
strength ($B$) and the swept-up mass ($M_{su}$) of interstellar
medium, respectively.}\label{fig:distrib}
\end{figure*}

There are many measures to obtain the age values ($t$) of remnants
(Xu et al. 2005). For example, if a remnant is associated with a
pulsar, we can estimate its age by using the neutron star
characteristic age obtained from the rotation period of the pulsar
(P) and the rate of change of period ($\dot{P}$) by $t=P/2\dot{P}$
(Gotthelf et al. 2000). For SNRs with a known radius ($R$) and
thermal temperature ($T$) taken from X-ray data, one can obtain the
age by $t=3.8\times10^2R_{\rm pc}(kT)^{-1/2}_{\rm keV}$~yr (Seward
et al. 1995). We can also calculate the SNR age by
$t~\approx~40000B^{-1.5}\nu_{\rm b}^{-0.5}$~yr, when a remnant has
its spectrum showing the usual break at frequency $\nu_b$ due to
synchrotron losses in a magnetic field $B$ (Bock et al. 2001).

In our statistics the age ($t$) (and also the Galactic height ($z$))
values are taken from the table in Xu et al. (2005).

\subsection{Derived Distances}

There are some different methods to derive the distance value ($d$)
of supernova remnants (Xu et al. 2005) in Galaxy. These already
known distances were derived by the optical proper motion/velocity,
HI absorption, association with CO/HI/HII region, HI column density,
pulsar parallax and optical velocity, or by various methods (e.g.,
Green 2004; Xu et al. 2005). Some remnants distance was estimated by
the radio surface brightness ($\Sigma$) and linear diameter $D$
relation. Obviously the former methods are somewhat more reliable.

\subsection{Initial Explosion Energy}

One first method to derive SNe initial energy, for example, by $E =
\frac{1}{2} M_{su}\upsilon^2$, here $M_{su}$ is the swept-up mass of
remnant shell expanding into interstellar media (ISM), $\upsilon$ is
the velocity of shock wave of remnant, the initial explosion energy
of SNR G180.0$-$1.7 is thus obtained (Braun et al. 1989). Sun et al.
(1999) calculating their detected ASCA data plus ROSAT data has
derived the initial energy of G327.1$-$1.1 as the fitting result.
After knowing the SNR G299.2$-$2.9 radius ($R$) value, the particle
density ($n_0$) and the age ($t$), Slane et al. (1996) get $E_0$
value by $E_0 \approx 340 R^5 n_0 t^{-2} \times 10^{51}$~ergs. Bamba
et al. (2001) have got the $E_0$ value by assuming a thin thermal
NEI plasma model plus standard Sedov model.

\subsection{ISM Density}

With emission measure (EM) Landecker et al. (1999) got $EM = \int
n_e ndV = 4.1\times10^{55} (D_{2.5})^2$~cm$^{-3}$ for remnant
G93.3+6.9, and then $n_0 = 5.9\times 10^{-30} (EM)^{1/2}
(D_{2.5})^{-3/2} \sim 0.01$~cm$^{-3}$. For SNR G82.2+5.3
Mavromatakis et al. (2004) calculated the hydrogen number density in
cm$^{-3}$ is $n_H = 0.657\times \sqrt{(EM)/d_{1.6}}$. Here $d_{1.6}$
is the distance to the remnant in unit of 1.6~kpc. In particular,
for SNR G291.0$-$0.1 Harrus et al. (1998) estimated the pre-shock
electron number density ($n_0$) through integrating the interior
radial density variation from the Sedov solution and then equating
it to the detected emission measure (EM). And so on.

\subsection{Hydrogen Column Density}

Methods to derive the hydrogen column density ($N_H$) are usually by
fitting the spectrum with different models: power law, thermal
bremsstrahlung, black body and thermal hot plasma, etc.. Sidoli et
al. (2000) found the best fit model for SNR G0.9+0.1 is a power-law
with photo index $\Gamma = 1.95$ and got the absorption column
density $N_H = 10^{23}$~cm$^{-2}$. For more others, one can see:
Craig et al. (1997) for SNR G116.9+0.2 of which $N_H = 7\times
10^{21}$~cm$^{-2}$, Kothes et al. (2001) for SNR G106.3+2.7 of which
$N_H = 6.3\times 10^{21}$~cm$^{-2}$, Byun et al. (2006) for
G89.0+4.7 of which $N_H = (3.5\pm 0.4)\times 10^{21}$~cm$^{-2}$.

\subsection{Electron Temperature}

For supernova remnant G352.7$-$0.1, Kinugasa et al. (1998) found
that the spectra are nicely fitted by a non-equilibrium ionozation
(NEI) model with an electron temperature $kT \sim 2.0$~KeV. Lazendic
et al. (2005) discover both soft and hard components for SNR
G349.7+0.2. The soft one is in ionization equilibrium and has a
temperature $T \approx 0.8$~KeV, the hard spectral one has a
temperature $T \approx 1.4$~KeV. But for remnant G49.2$-$0.7 Koo et
al. (2005) fitted the overall spectral shapes reasonably well with a
single component equilibrium thermal emission model at temperature
of $kT = 0.3 -0.5$~KeV.

\subsection{Magnetic Field}

The remnants magnetic field ($B$) could be calculated by rotation
measure value (RM) plus Galactic electron density distribution. For
example, Gaensler et al. (1998) obtain the mean ISM magnetic field
along the line of sight in the range $1 - 9~\mu G$. The second one
is using the Zeeman effect (Brogan et al. 2000). And there are also
many some other special measures (Ruiz \& May 1986; Sidoli et al.
2000; Gaensler et al. 2002).

\subsection{Swept-up ISM Mass}

For example, to a middle-age remnant in the uniform interstellar
medium and through the analytical Sedov solution, for SNR G
337.2$-$0.7 Rakowski et al. (2006) got
\begin{eqnarray}
M_{su} = 39.7 \left( \begin{array}{c} \frac{N}{2.0\times 10^{12}
cm^{-5}} \end{array} \right) ^{1/2}\nonumber\\
\times \left( \begin{array}{c} \frac{\theta_R}{2'.75} \end{array}
\right) ^{3/2} \left( \begin{array}{c} \frac{D}{9.6kpc}
\end{array} \right) ^{5/2} M\odot
\end{eqnarray}
Here, the normalization $N = n_e n_H V/4\pi D^2$, $\theta_R$ is the
maximum angular radius, $D$ is an upper limit on the remnant
distance.

\section{Statistical Results}
\label{sec:statis}

For the image of the swept-up mass and evolving age relation we took
samples of 18 shell-type and total 29 SNRs (Fig.~\ref{fig:rela}),
the electrons temperature and age relation samples of 30 shell-type
and total 43 SNRs, the waves traveling velocity and age relation
samples of 41 shell-type and total 59 SNRs, the radio kluxes and age
relation samples of 67 shell-type and total 95 SNRs and at last the
remnants magnetic field and age one of 15 shell-type and total 27
SNRs.

\subsection{SNR Parameters versus Age}
\label{sec:age}

Several parameters of the supernova remnant own significant
correlation with its evolving age ($t$). For example, the electrons
temperature ($T$) of hard X-ray and the shock waves traveling
velocity ($\upsilon$) decreases with ages for all-sort remnants,
with the slope of best fitting line $-$0.320 and $-$0.491,
respectively (Fig.~\ref{fig:rela}). However the shell swept-up mass
($M_{su}$) increases with the age, with a fit line slope of about
0.660. Although these plots dispersion is somewhat large, it seems
that there are detectable synchrotron emission energy loss of the
non-thermal electrons in shock waves, leading to the practically
decreases of the hard X-ray electrons temperature and the remnants
traveling velocity.
\begin{equation}
M_{su} = 0.171\times t_{yr}^{0.660\pm0.124}~(M_{\odot})
\end{equation}
\begin{equation}
\upsilon = 4.17\times 10^4 t_{yr}^{-0.491\pm0.085}~(km~s^{-1})
\end{equation}
\begin{equation}
T = 1.50\times 10^7 t_{yr}^{-0.320\pm0.063}~(KeV)
\end{equation}

It shows no obvious meaningful correlation between the SNRs magnetic
field ($B$) and evolved age ($t$). This is most likely because that
the detected magnetic field denotes those along the line of sight
and not mainly the one inside the remnant. Therefore the magnetic
field almost remain unchange with time.

The plot also shows an unchangeable radio fluxes of SNRs at 1~GHz
(middle-right in Fig.~\ref{fig:rela}). Although there are energy
loss of the synchrotron emission of the non-thermal electrons, this
has no or not remarkable influence on the remnants radio fluxes.

\subsection{Radio Fluxes against ISM Density}
\label{sec:density}

For the image of the remnants radio fluxes ($S_{1GHz}$) and electron
density ($n_0$) relation we took samples of 54 shell-type and total
69 SNRs (lower-right in Fig.~\ref{fig:rela}). It tends that when the
particle density ($n_0$) of the interstellar medium increases, then
to some extent the SNRs radio fluxes would increase slightly but not
very remarkably, since the dispersion in the plot is somewhat large.
\begin{equation}
S_{1GHz} = 19.4\times n_{0,cm^{-3}}^{0.163\pm0.047}~(Jy)
\end{equation}

It is another evidence that the SNRs (for example, SNR G156.2+5.7)
appear enhanced radio emission towards a denser interstellar
environment (Xu et al. 2007). When the remnant shock waves interact
with a thinner cloud gases the radio radiations will be weaker, and
even disappear if lack of surrounding media as the shock waves
travel along.

\subsection{Number Distributions}
\label{sec:distrib}

We plot the SNRs number distributions of the electron density
($n_0$), SNe initial kinetic energy ($E_0$), hydrogen column density
($N_H$), electron temperature (kT) of hard X-ray, magnetic field
($B$) and the shock waves swept-up mass ($M_{su}$) of ISM in
Fig.~\ref{fig:distrib} for shell-type and other type SNRs,
respectively. We found that the supernovae initial explosion energy
peaked at about $10^{50} \sim 10^{51}$~ergs, just as the publicly
accepted value. The typical value of the hydrogen column density
equals about to $(1\sim 10)\times 10^{21}$~cm$^{-2}$. SNRs magnetic
field range from 1~$\mu$G to $10^4$~$\mu$G, mainly concentrates on
$\sim 100$~$\mu$G. Most remnants with the ISM swept-up mass about
$10\sim 100$~$M_{\odot}$ were detected. The swept-up mass of an SNR
could reach as high as $10^5$~$M_{\odot}$.

At last, there are no other rather meaningful correlations among all
these SNRs parameters (table~\ref{tab1}) could be discovered.

\section{Summary}
\label{summary}

Our statistics on Galactic supernova remnants (SNRs) shows that
\begin{enumerate}
\item The hard X-ray electrons temperature and the shock waves
traveling velocity decreases with ages for the all-type remnants.
\item  The shock waves swept-up mass of ISM increases with
the age.
\item Remnants magnetic field and radio fluxes almost remain a
constant with time.
\item The remnants radio fluxes at 1~GHz increase with ISM electrons
density.
\item The supernova initial energy, hydrogen column density, electron
temperature, magnetic field and the shock waves swept-up mass of ISM
has a peak-shape distribution.
\end{enumerate}

\section*{Acknowledgments}

JWX thanks Y. P. Hu for his useful assistance and help during the
paper work.

\appendix

\begin{table*}
\caption{Some physical parameters of Galactic supernova remnants
adopted on our statistics.} \label{tab1}
\begin{tabular}{@{}llllllllllll}
  \hline
  \hline
   Source    & Dia./ &$\Upsilon$/& $N_0$/   & $E_0$/        & N$_H$/        & kT/        & B/     & M$_{su}$/  & Ref.\\
   $-$       & pc    &Km~        & cm$^{-3}$&$\times10^{51}$&$\times10^{21}$& KeV        & $\mu$G & M$_{\odot}$&\\
   $-$       &       &s$^{-1}$   &          & ~ergs         & $cm^{-2}$     &            &        &            &\\
  \hline
   G0.0$+$0.0 & 7   & ---     & 10000    & $\geq40$   & 150         &  1/5       & 3000?    & ---     & 05PMB,96KF,98KFG\\
   G0.3$+$0.0 & 17  & 100     & ---      & ---        & ---         &  ---       & 70       & ---     & 96KF  \\
   G0.9$+$0.1 & 23  & 10000   & ---      & ---        & 100         &  ---       & 67       & ---     & 00SMI\\
   G4.5$+$6.8 & 3   & 5300    & 0.5      & 0.4/0.5    & ---         &  3.1       & 113      & 1.4/5   & 05BKV,06V,89BGL\\
   G5.4$-$1.2 & 47  & 240?    & ---      & ---        & ---         &  ---       & 5-10     & ---     & 06BGC,89SFS\\
   G6.4$-$0.1 & 26  & 170?    & ---      & ---        & 5.25        &  0.62,1.39 & 750      & ---     & 05HGB,89SFS,05KON\\
   G7.7$-$3.7 & 29  & ---     & ---      & ---        & ---         &  ---       & ---      & ---     & 86MRK\\
   G8.7$-$0.1 & 51  & ---     & ---      & ---        & ---         &  ---       & ---      & ---     & 94FKW\\
   G9.8$+$0.6 & 43  & ---     & ---      & ---        & ---         &  ---       & ---      & ---     & 83C\\
  G11.2$-$0.3 & 5   & ---     & ---      & ---        & 22.2        &  ---       & ---      & 3-5     & 03TR,06KM,05C\\
  G13.3$-$1.3 & 50  & $\leq$60& ---      & ---        & 5.8         &  ---       & ---      & ---     & 95SDF\\
  G15.9$+$0.2 & 29  & 1000    & 0.7      & ---        & 40          &  0.9$\pm$0.1& ---     & 22      & 06RBH\\
  G16.2$-$2.7 & 43  & ---     & ---      & ---        & ---         &  ---       & ---      & ---     & 99T\\
  G16.7$+$0.1 & 2   & 3000    & ---      & ---        & 47.4        &  ---       & ---      & ---     & 03HAG\\
  G17.4$-$2.3 & 44  & 100     & 0.1      & ---        & 2-5         &  1?        & ---      & ---     & 02BMP\\
  G18.8$+$0.3 & 57  & $\geq$10& 600      & 0.01       & 19          &  ---       & ---      & 1100    & 99DGR\\
  G18.9$-$1.1 & --- & ---     & 0.06     & 20         & 8.3-9.4     &  0.58-1.12 & ---      & ---     & 04HSH, 97DSC\\
  G20.0$-$0.2 & 12  & ---     & ---      & ---        & ---         &  ---       & ---      & ---     & 85BH\\
  G21.5$-$0.9 & 5   & 3200?   & ---      & ---        & 20          &  ---       & 300?     & 10?     & 05KLS, 06CRG\\
  G21.8$-$0.6 & 17  & ---     & ---      & ---        & ---         &  ---       & 43       & ---     & 75VK\\
  G24.7$+$0.6 & 32  & 670     & 0.018?   & 0.066      & ---         &  0.54      & ---      & ---     & 87BH,89L\\
  G27.4$+$0.0 & 8   & ---     & ---      & ---        & 16-23       &  ---       & ---      & ---     & 97VG, 01BUK\\
  G28.6$-$0.1 & --- & ---     & 0.2      & 0.9        & 24-40       &  5.4       & ---      & 20      & 01BUK\\
  G29.6$+$0.1 & 16  & ---     & ---      & ---        & 100         &  ---       & ---      & ---     & 00VGT,04AH\\
  G29.7$-$0.3 & 17  & 3700    & 0.1      & 2          & 39.6        &  2.99      & ---      & 5       & 03HCG,05C  \\
  G31.9$+$0.0 & 13  & $\leq$30& 1.1      & 0.3-1.4    & 2.7-4.1     &  0.46-0.79 & ---      & ---     & 05CSS,03YMR,03K\\
  G32.1$-$0.9 & 64  & ---     & 0.05     & ---        & 2.3         &  0.8       & ---      & ---     & 97FWW\\
  G32.8$-$0.1 & 35  & ---     & ---      & ---        & ---         &  ---       & 1450?    & ---     & 98KFG\\
  G33.2$-$0.6 & 38  & 50      & ---      & ---        & ---         &  ---       & 85-180   & ---     & 82R\\
  G34.7$-$0.4 & 25  & $\leq$20& ---      & ---        & 9.6         &  1         & 200?     & ---     & 98KFG,05KON,03K\\
  G39.7$-$2.0 & 73  &$\leq$4000?& 0.01   & 1          & ---         &  ---       & $\geq$10?& ---     & 96MS   \\
  G42.8$+$0.6 & 42  & ---     & ---      & ---        & ---         &  ---       & ---      & ---     & 03CSC\\
  G43.3$-$0.2 & 10  & ---     & 3        & ---        & ---         &  2.2-2.7   & ---      & ---     & 01LLK,01BUK,89BGL\\
  G49.2$-$0.7 & 52  & 120?    & 50-100   & ---        & 17-22       &  0.3-0.38  & 100      & ---     & 05KLS\\
  G53.6$-$2.2 & 24  & 60      & 4-21     & ---        & ---         &  ---       & ---      & ---     & 06ARF\\
  G54.1$+$0.3 & 4   & ---     & 0.2-0.3  & ---        & 16          &  ---       & ---      & ---     & 02CLB,06KM,89L,05C\\
  G54.4$-$0.3 & 38  & 70-80   & $\leq$50 & ---        & 10-40       &  2         & ---      & 50000   & 92JFRa,b,05BMX\\
  G55.0$+$0.3 & 71  & ---     & 0.8      & ---        & ---         &  ---       & ---      & ---     & 98MWT\\
  G59.8$+$1.2 & 48  & $\leq$70& 130      & ---        & 26          &  ---       & ---      & ---     & 05BMX,89SRH\\
  G63.7$+$1.1 & --- & ---     & ---      & ---        & 0.044?      &  ---       & ---      &51$\pm$43& 97WLT\\
  G65.3$+$5.7 & 78  & 155?    & 0.02-200 & ---        & 1.4         &  ---       & ---      & ---     & 04BML,06KGK\\
  G69.0$+$2.7 & 46  & 750     & ---      & ---        & 6           &  ---       & 23       & ---     & 05KLS\\
  G73.9$+$0.9 & 8   & 30-50?  & 5.4?     & ---        & ---         &  ---       & ---      & 860?    & 96PGM  \\
  G74.0$-$8.5 & 23  & 180     & 10       & 0.24       & 0.8         &  0.3       & ---      & ---     & 05BSR,06KGK,89BGL\\
  G74.9$+$1.2 & 12  & ---     & ---      & ---        & ---         &  ---       & ---      & ---     & 03KRF\\
  G78.2$+$2.1 & 26  & 1000    & 25       & ---        & >5          &  1.5       & ---      & ---     & 04BKU,89BGL\\
  G82.2$+$5.3 & 36  & 100     & 0.04     & 0.17       & 4           &  0.2       & ---      & ---     & 04MAB \\
  G84.2$-$0.8 & 23  & ---     & 300      & ---        & ---         &  ---       & ---      & ---     & 93FG\\
  G89.0$+$4.7 & 24  & 45?     & 0.18     & 3.5$\pm$0.4& 0.2$\pm$0.03& ---        & ---      & 390     & 06BKT\\
  \hline
\end{tabular}
\end{table*}

\begin{table*}
\begin{tabular}{@{}llllllllllll}
  \hline
  \hline
   Source    & Dia./&$\Upsilon$/& $N_0$/   & $E_0$/        & N$_H$/        & kT/      & B/     & M$_{su}$/  & Ref.\\
   $-$       & pc   &Km~        & cm$^{-3}$&$\times10^{51}$&$\times10^{21}$& KeV      & $\mu$G & M$_{\odot}$&\\
   $-$       &      &s$^{-1}$   &          & ~ergs         & $cm^{-2}$     &          &        &            &\\
  \hline
  G93.3$+$6.9 & 15  &$\geq$3200& ---     & 0.39       & 5.7         &  12        & 2.3      & 3.9     & 99LRR,04GAT\\
  G93.7$-$0.2 & 35  & ---     & 0.01     & ---        & ---         &  ---       & 0.5      & ---     & 02UKB    \\
  G94.0$+$1.0 & 37  & 48      & 0.2?     & $\geq$0.27 & 16          &  0.45      & 15       & 55      & 05F\\
 G106.3$+$2.7 & 139 & ---     & 100      & 0.07       & 2           &  ---       & ---      & 40?     & 01KUP\\
 G109.1$-$1.0 & 24  & ---     & 0.25?    & 1-10       & 4.5         &  0.9       & ---      & ---     & 06SKP,89BGL,04GAT\\
 G111.7$-$2.1 & 5   & 5200    & 1        & 2-3        & 1.8         &  0.5       & 299?     & ---     & 06FPS,80CD,06V,89BGL\\
 G114.3$+$0.3 & 15  & ---     & 0.03     & ---        & 2           &  ---       & ---      & ---     & 96BBT,98MWT,04GAT\\
 G116.5$+$1.1 & 32  & 70-120  & ---      & ---        & 5           &  ---       & ---      & ---     & 05MBX\\
 G116.9$+$0.2 & 16  & 505     & 0.08     & 0.1        & 7           &  0.3       & ---      & ---     & 97CHP\\
 G119.5$+$10.2& 37  & 400     & 0.02-1   & 0.03       & 3.8         &  0.16-1.14 & 2.9      & 13      & 04TDR,04GAT\\
 G120.1$+$1.4 & 5   & 3400?   & 0.3      & 1.16       & 6.4         &  2.15      & 16.5     & ---     & 05WHB,75VK,06V\\
 G126.2$+$1.6 & 91  & 100     & 13.3     & 7          & 8.0         &  0.05?     & ---      & ---     & 79RKS,05BMX\\
 G127.1$+$0.5 & 69  & 95?     & ---      & ---        & ---         &  ---       & ---      & ---     & 84GG,04GAT\\
 G130.7$+$3.1 & 6   & 4000    & 1?       & ---        & 1.8-3       &  ---       & 80?      & 43      & 06CRG,89BGL,04GAT\\
 G132.7$+$1.3 & 51  & 20      & 0.27     & 0.31       & 3-6.9       &  0.3,1     & ---      & 500     & 91RDL,89BGL,04GAT\\
 G156.2$+$5.7 & 64  & ---     & ---      & ---        & 3.5         &  ---       & ---      & ---     & 06KGK,07XHS\\
 G160.9$+$2.6 & 38  & ---     & 10-100   & ---        & 1           &  0.3       & ---      & ---     & 98MWT,06KGK,89BGL\\
 G166.0$+$4.3 & 57  &$\geq$100& 100      & ---        & 2.9         &  0.83      & ---      & ---     & 97GB,89BGL \\
 G179.0$+$2.6 & 59  & ---     & ---      & ---        & ---         &  ---       & 3-4      & ---     & 89FRK,04GAT\\
 G180.0$-$1.7 & 84  & 100     & 10       & ---        & ---         &  ---       & ---      & ---     & 96ACJ,89BGL,04GAT\\
 G182.4$+$4.3 & 44  & 2300    & 0.013    & 0.24       & $\leq$4     &  5         & ---      & 14      & 98KFR\\
 G184.6$-$5.8 & 3   & 300?    & ---      & 0.015      & 3           &  ---       & ---      & 2       & 89SRH,06KM,04GAT\\
 G189.1$+$3.0 & 20  & 450?    & 100?     & ---        & ---         &  1.2       & 42       & 30      & 05KLS,75VK,89BGL\\
 G192.8$-$1.1 & 59  & ---     & ---      & ---        & ---         &  ---       & ---      & ---     & 85C \\
 G205.5$+$0.5 & 102 & 45      & 0.4      & ---        & $\leq$0.8   &  ---       & ---      & ---     & 86O,06KGK\\
 G206.9$+$2.3 & 102 & 400     & 0.01?    & 0.75       & 1           &  0.14      & ---      & ---     & 86L   \\
 G260.4$-$3.4 & 35  & 650     & 0.4-1    & ---        & 2.9-4.7     &  0.3       & ---      & ---     & 95BRL,89BGL,04GAT\\
 G261.9$+$5.5 & 40  & $\leq$14& 0.23     & 0.29       & 0.73        &  ---       & ---      & 10000   & 80CD\\
 G263.9$-$3.3 & 22  & 100-140 & 10-100   & 1-2        & 0.3         &  0.01-0.1  & 50-85    & ---     & 06NEK,89BGL,04GAT\\
 G266.2$-$1.2 & 52  & 15000   & 1        & ---        & 2.6-4.1     &  4.4/6.5   & ---      & ---     & 05KBL,05IAB \\
 G272.2$-$3.2 & 8   & ---     & ---      & ---        & ---         &  ---       & ---      & ---     & 97DSC\\
 G279.0$+$1.1 & 83  & ---     & ---      & ---        & ---         &  ---       & ---      & ---     & 95DHS\\
 G284.3$-$1.8 & 20  & 16.9?   & 10       & ---        & ---         &  ---       & 1        & ---     & 86RM  \\
 G290.1$-$0.8 & 28  & ---     & 1        & 0.8        & 13          &  0.6       & ---      & ---     & 02SSH\\
 G291.0$-$0.1 & 41  & ---     & 0.034    & 0.25?      & 3?          &  0.8       & ---      & 30?     & 98HHSb\\
 G292.0$+$1.8 & 17  & 1700    & 1        & 0.18       & ---         &  0.4       & ---      & 20      & 05GHW,89BGL\\
 G292.2$-$0.5 & 43  & ---     & ---      & 1.0        & 5           &  4         & 100-280  & ---     & 05GS,05C\\
 G296.1$-$0.5 & 69  & ---     & 0.8      & 0.23       & 1.0         &  0.2       & ---      & 250     & 94HM\\
 G296.5$+$10.0& 44  & ---     & ---      & ---        & 1.0-1.8     &  0.17      & ---      & ---     & 98ZPT \\
 G296.8$-$0.3 & 47  & 6500    & 0.2?     & 0.2-0.6    & ---         &  ---       & 1.7      & 100     & 98GMG  \\
 G299.2$-$2.9 & 2   & 700     & 0.093    & 0.12       & 0.3         &  0.6       & ---      & ---     & 96SVH\\
 G308.8$-$0.1 & 50  & ---     & ---      & ---        & ---         &  ---       & ---      & ---     & 92KMJ\\
 G309.2$-$0.6 & 16  & ---     & 0.2      & ---        & 10          &  ---       & 1-9?     & 20-170  & 98GGM\\
 G312.4$-$0.4 & 34  & ---     & 0.8      & 0.6        & ---         &  ---       & 6        & ---     & 99CB\\
 G315.4$-$2.3 & 28  & 3500?   & 100?     & 0.66       & 4.3-7.0     &  1.4       & 10       & ---     & 03WTK,89BGL,05BYY\\
 G315.4$-$0.3 & 37  & ---     & ---      & ---        & ---         &  ---       & ---      & ---     & 81CMW\\
 G318.2$+$0.1 & --- & ---     & 1?       & ---        & 15          &  5?        & ---      & ---     & 01BPM\\
 G320.4$-$1.2 & 53  & ---     & 100      & 1-2        & 9.5         &  1.3       & ---      & ---     & 02GAP,06KM,89BGL\\
 G321.9$-$0.3 & 70  & 340     & ---      & ---        & ---         &  ---       & ---      & ---     & 02MLC,89SFS\\
  \hline
\end{tabular}
\end{table*}

\begin{table*}
\begin{minipage}{140mm}
\begin{tabular}{@{}llllllllllll}
 \hline
 \hline
   Source    &  Dia./ &$\Upsilon$/& $N_0$/   & $E_0$/        & N$_H$/        & kT/      & B/     & M$_{su}$/  & Ref.\\
   $-$       &  pc    &Km~        & cm$^{-3}$&$\times10^{51}$&$\times10^{21}$& KeV      & $\mu$G & M$_{\odot}$&\\
   $-$       &        &s$^{-1}$   &          & ~ergs         & $cm^{-2}$     &          &        &            &\\
 \hline
 G322.5$-$0.1 & 45  & ---     & ---      & ---        & ---         &  ---       & ---      & ---     & 92W\\
 G326.3$-$1.8 & 41  & ---     & ---      & 1.0        & 8.9         &  ---       & 45       & ---     & 99SWC, 00DMS\\
 G327.1$-$1.1 & 46  & 600     & 0.1      & 0.23       & 18          &  0.37      & 65       & 49      & 99SWC  \\
 G327.4$+$0.4 & 29  & ---     & 0.39     & 0.14       & 22          &  0.71      & ---      & 34      & 02ETM,05KON \\
 G327.6$+$14.6& 19  & 5160?   & 0.1      & 1.0        & ---         &  0.1,1     & 39?      & ---     & 05WLH,06V,89BGL\\
 G328.4$+$0.2 & 25  & ---     & ---      & 0.9-1.8    & 100         &  4.0       & 220      & ---     & 00HSP   \\
 G330.0$+$15.0& 63  & ---     & ---      & 0.005-0.008& 0.5         &  0.18-0.32 & ---      & ---     & 90GGL,06KGK\\
 G330.2$+$1.0 & 33  & ---     & ---      & ---        & ---         &  ---       & ---      & ---     & 83CHMb\\
 G332.4$-$0.4 & 9   & 10-20   & 100?     & ---        & ---         &  0.5       & ---      & 4000    & 06PRP,89BGL\\
 G332.4$+$0.1 & 22  & 5000?   & ---      & ---        & 40          &  1         & ---      & ---     & 04V\\
 G337.0$-$0.1 & 5   & ---     & ---      & ---        & ---         &  ---       & 1100?    & ---     & 00BFG,04AH\\
 G337.2$-$0.7 & 26  & ---     & ---      & ---        & 32-35       &  0.74-0.85 & ---      & ---     & 06RBG\\
 G337.3$+$1.0 & 22  & ---     & ---      & ---        & ---         &  ---       & ---      & ---     & 89MCK\\
 G337.8$-$0.1 & 27  & ---     & ---      & ---        & ---         &  ---       & ---      & ---     & 98KFG  \\
 G340.4$+$0.4 & 36  & ---     & ---      & ---        & ---         &  ---       & ---      & ---     & 83CHMa\\
 G340.6$+$0.3 & 29  & ---     & ---      & ---        & ---         &  ---       & ---      & ---     & 83CHMa\\
 G341.9$-$0.3 & 35  & ---     & ---      & ---        & ---         &  ---       & ---      & ---     & 83CHMa\\
 G342.0$-$0.2 & 50  & ---     & ---      & ---        & ---         &  ---       & ---      & ---     & 83CHMa\\
 G343.1$-$2.3 & 19  &$\leq$500& 0.00023? & ---        & 2-5         &  ---       & ---      & ---     & 02BG\\
 G346.6$-$0.2 & 19  & ---     & ---      & ---        & ---         &  ---       & 1700?    & ---     & 98KFG,04AH\\
 G347.3$-$0.5 & 104 & 5500    & 0.01     & 0.02       & ---         &  35        & ---      & $\leq$3 & 05MTT\\
 G348.5$-$0.0 & 21  & ---     & ---      & ---        & ---         &  ---       & 850?     & ---     & 91KBW,00BFG\\
 G349.7$+$0.2 & 9   & 710     & 5-25     & 0.5        & 70          &  0.8/1.4   & 300      & 160     & 05LSH,00BFG\\
 G351.2$+$0.1 & 22  & ---     & ---      & ---        & ---         &  ---       & ---      & ---     & 83CHMc\\
 G352.7$-$0.1 & 17  & ---     & 0.1      & 0.2        & 29          &  2.0       & ---      & ---     & 98KTT\\
 G355.9$-$2.5 & 30  & ---     & ---      & ---        & ---         &  ---       & ---      & ---     & 83CHMc\\
 G357.7$-$0.1 & 7   & 140?    & ---      & ---        & ---         &  ---       & ---      & 15000   & 88GK,89SFS\\
 G357.7$+$0.3 & 45  & 670     & 0.025    & 0.11       & ---         &  ---       & ---      & ---     & 99YGR,89L\\
 G359.0$-$0.9 & 33  & 670     & 0.055    & 0.21       & ---         &  ---       & ---      & ---     & 89L\\
 G359.1$-$0.5 & 37  & ---     & ---      & ---        & ---         &  ---       & ---      & ---     & 03YMR\\
 \hline
\end{tabular}
\end{minipage}
\end{table*}

\label{lastpage}

\end{document}